\documentclass{article}
\usepackage{graphicx}
\usepackage{amssymb}
\begin{document}
\title{\textbf{An approach to construct wave packets with complete classical-quantum correspondence in non-relativistic quantum mechanics}}
\author{Pouria Pedram\thanks{pouria.pedram@gmail.com}\\
{\small Plasma Physics Research Center, Science and Research Campus,
Islamic Azad University, Tehran, Iran}}

\date{\today}
\maketitle \baselineskip 24pt
\begin{abstract}
We introduce a method to construct wave packets with complete
classical and quantum correspondence in one-dimensional
non-relativistic quantum mechanics. First, we consider two similar
oscillators with equal total energy. In classical domain, we can
easily solve this model and obtain the trajectories in the space of
variables. This picture in the quantum level is equivalent with a
hyperbolic partial differential equation which gives us a freedom
for choosing the initial wave function and its initial slope. By
taking advantage of this freedom, we propose a method to choose an
appropriate initial condition which is independent from the form of
the oscillators. We then construct the wave packets for some cases
and show that these wave packets closely follow the whole classical
trajectories and peak on them. Moreover,  we use de-Broglie Bohm
interpretation of quantum mechanics to quantify this correspondence
and show that the resulting Bohmian trajectories are also in a
complete agreement with their classical counterparts.
\end{abstract}



\section{Introduction}
The issue of classical-quantum correspondence has been extensively
investigated in the literature \cite{Bolivar}. Moreover, the
question of construction and interpretation of wave packets in
quantum mechanics and its connection with classical mechanics has
been attracting much attention. In quantum physics, one is generally
concerned with the construction of wave packets by the superposition
of the energy eigenstates which would peak around the classical
trajectories. These efforts have been started by Schr\"odinger
\cite{Ervin} and followed by others who were interested in finding
quantum mechanical states which provide a close connection between
classical and quantum formulations of a given physical system in the
context of the coherent states \cite{ref}. These states can also be
generated using algebraic methods \cite{algebra}, supersymmetric
quantum mechanics \cite{susy} and its application to different
physical situations \cite{ref2}.

In this paper, we pursue a different approach to construct wave
packets with complete classical and quantum correspondence. First,
we consider similar Schr\"odinger equations of two oscillators with
$u$ and $v$ variables. Classically, we can easily solve this model
and find the behavior of the variables versus time ($u(t),v(t)$). If
these two oscillators have the same total energy, the solutions will
be equivalent up to a temporal phase factor. Imposing particular
initial conditions, we will obtain the trajectories in the
configuration space ($u,v$). Inversely, using the form of the
trajectories, again we can find the temporal behavior of the
variables. For instance, in general, the trajectory for the case of
a simple harmonic oscillator (SHO) is an ellipse. Therefore, we can
obtain the time behavior of the variables as $u(t)=A\cos(\omega t)$
and $v(t)=A\sin(\omega t+\phi_0)$, where $\phi_0=0$ corresponds to a
circle. Thus, we can parameterize the behavior of one variable
(\textit{e.g.} $u$) versus another variable (\textit{e.g.} $v$)
where they both satisfy the same equation of motion.

Our motivation for this model is related to its interesting quantum
mechanical properties. In the quantum version, this model allows us
to define a wave function which is a function of $u$ and $v$ unlike
the usual case where the wave function is a function of $u$ or $v$
and time. As we shall see, the existence of the two mentioned
similar oscillators results in a hyperbolic partial differential
equation (PDE). Therefore, we are free to choose the initial wave
function and the initial slope of the wave function. This freedom
for choosing the initial conditions (or the expansion coefficients)
allows us to obtain a specific form of the wave packets with
complete classical and quantum correspondence. In fact, the initial
wave function and its derivative correspond to classical initial
position and initial momentum, respectively. To fix the expansion
coefficients, we will apply the prescription used in
Refs.~\cite{pedram,pedramPLB3,pedramJCAP} for the case of hyperbolic
PDEs which also appear in the context of quantum cosmology.

The purpose of this paper is to construct wave packets such that the
classical and quantum correspondence is manifest, which means that
the wave packet should be centered around the classical path and the
crest of the wave packet should follow the classical path as closely
as possible. Usually, the temporal behavior of the variables is
studied in literature where its underlying equation is parabolic.
But here, we are encountered with the hyperbolic PDEs which
correspond to the classical picture of parametric trajectories.
Although both descriptions are equivalent, the latter contains some
interesting properties in quantum domain which allows us to obtain
the complete classical-quantum correspondence for any oscillator
using a unique prescription.

To be more precise, we can also quantify this correspondence using
de-Broglie Bohm interpretation of quantum mechanics \cite{holland}.
This approach gives us the Bohmian trajectories which are guided by
the wave function. These trajectories are governed by classical plus
quantum potential. Therefore, the coincidence between classical and
quantum trajectories accrues in the limit of vanishing quantum
potential. In fact, the usage of the appropriate initial conditions
results in the suppression of the quantum potential
\cite{pedramJCAP}.

The paper is organized as follows: In Sec.~\ref{sec2}, we present
the model of two similar oscillators which results in a hyperbolic
PDE at the quantum level. We then construct wave packets by choosing
the appropriate expansion coefficients which leads to a good
correspondence between classical and quantum solutions. In
Sec.~\ref{sec3}, we use the casual interpretation of quantum
mechanics to quantify this correspondence and obtain the Bohmian
trajectories for different cases. In Sec.~\ref{sec4}, we state our
conclusions.

\section{The model}\label{sec2}
In non-relativistic quantum mechanics we can write the dimensionless
time independent Schr\"odinger equation for variable $u$ as
\begin{equation}\label{Sch-1}
-\frac{d^2\psi(u)}{du^2}+V(u)\, \psi(u)=E\, \psi(u),
\end{equation}
where $V$ is the potential term which here is supposed to be an even
function of its variable and $E$ is the total energy. Also we can
rewrite the above equation for another variable $v$
\begin{equation}\label{Sch-2}
-\frac{d^2\psi(v)}{dv^2}+V(v)\, \psi(v)=E\, \psi(v).
\end{equation}
Classically, the above equations show two similar oscillatory
motions for $u$ and $v$ variables with equal total energy and the
following equations of motion
\begin{eqnarray}
\ddot{u}= -2\frac{dV(u)}{du},\hspace{2cm}\ddot{v}=
-2\frac{dV(v)}{dv}.
\end{eqnarray}
By assuming specific classical initial conditions, we can obtain a
particular trajectory in the configuration space. Here, we impose
the following initial conditions:
\begin{eqnarray}\label{class-initial}
v(0)= 0,\hspace{1cm}u(0)= u_0,\hspace{1cm}\dot{v}(0)=
\dot{v}_0,\hspace{1cm}\dot{u}(0)= 0,
\end{eqnarray}
which result in circular motion for the case of simple harmonic
oscillators for $u$ and $v$ variables (Fig.~\ref{fig1}). Moreover,
the classical behavior of $u$ versus time in the presence of some
sample potentials is also depicted in the right part of
Figs.~\ref{fig2}-\ref{fig5} as dashed lines. Note that, the above
initial conditions result in classical solutions which have the
following property
\begin{eqnarray}
v(t)=u(t-T/4),
\end{eqnarray}
where $T$ is the classical period of motion.

Now, we define a new wave function $\Psi(u,v)=\psi(u)\psi(v)$ where
$\psi(u)$  and  $\psi(v)$ satisfy Eqs.~(\ref{Sch-1}) and
(\ref{Sch-2}), respectively. Therefore, $\Psi(u,v)$ satisfies the
following partial differential equations
\begin{eqnarray}
\left\{
  \begin{array}{l}
   -\frac{\displaystyle\partial^2\Psi(u,v)}{\displaystyle\partial u^2}+V(u)\,
\Psi(u,v)=E\,
\Psi(u,v),\\
-\frac{\displaystyle\partial^2\Psi(u,v)}{\displaystyle\partial
v^2}+V(v)\, \Psi(u,v)=E\, \Psi(u,v).
\end{array}
\right.
\end{eqnarray}
Subtracting these equations leads to
\begin{equation}
\label{wd}\left\{- \frac{\partial^2}{\partial
u^2}+\frac{\partial^2}{\partial v^2}+V(u)-V(v)\right\}\,
\Psi(u,v)=0,
\end{equation}
where is a hyperbolic differential equation. This equation can be
solved using the separation of variables and the general wave packet
which satisfies this equation can be written as
\begin{equation}\label{psi}
\Psi(u,v)=\sum_{n=\mbox{\footnotesize{even}}} A_n
\psi_n(u)\psi_n(v)+i\sum_{n=\mbox{\footnotesize{odd}}} B_n
\psi_n(u)\psi_n(v).
\end{equation}
Since the potential is an even function of its variable, the
eigenfunctions are separated into even and odd categories. Moreover,
we have intentionally separated the odd and even terms for further
usages. To find the exact form of the wave function we need to
specify the expansion coefficients. These coefficients will be
determined from the initial form of the wave function which is
evaluated along $u-$axis ($v=0$) in consistency with the classical
initial conditions (\ref{class-initial}).

\begin{figure}
\centerline{\begin{tabular}{ccccc}
\includegraphics[width=5cm]{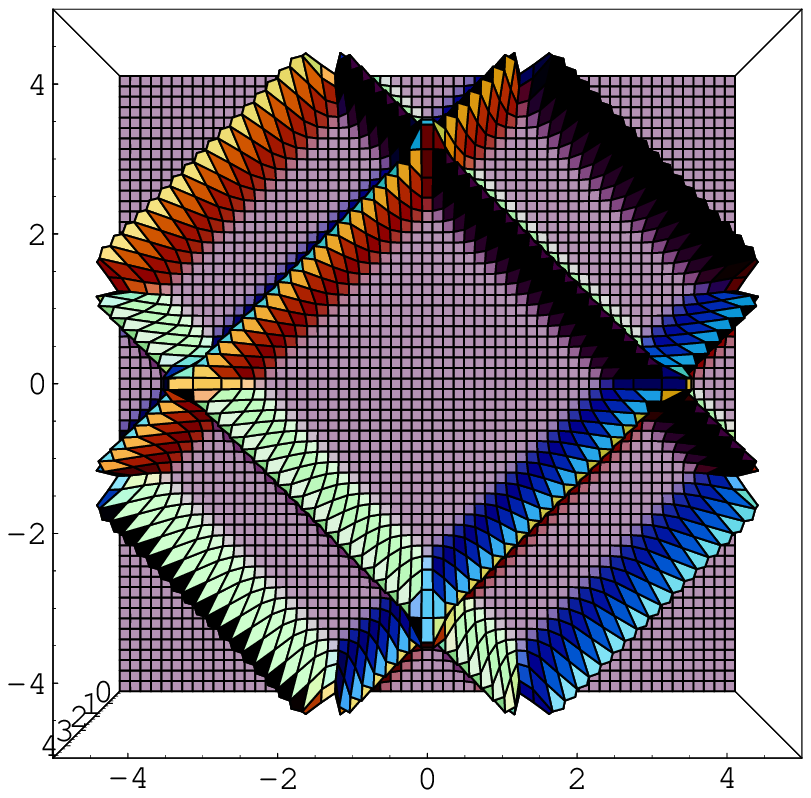}
 &\hspace{0.cm}&
\includegraphics[width=5cm]{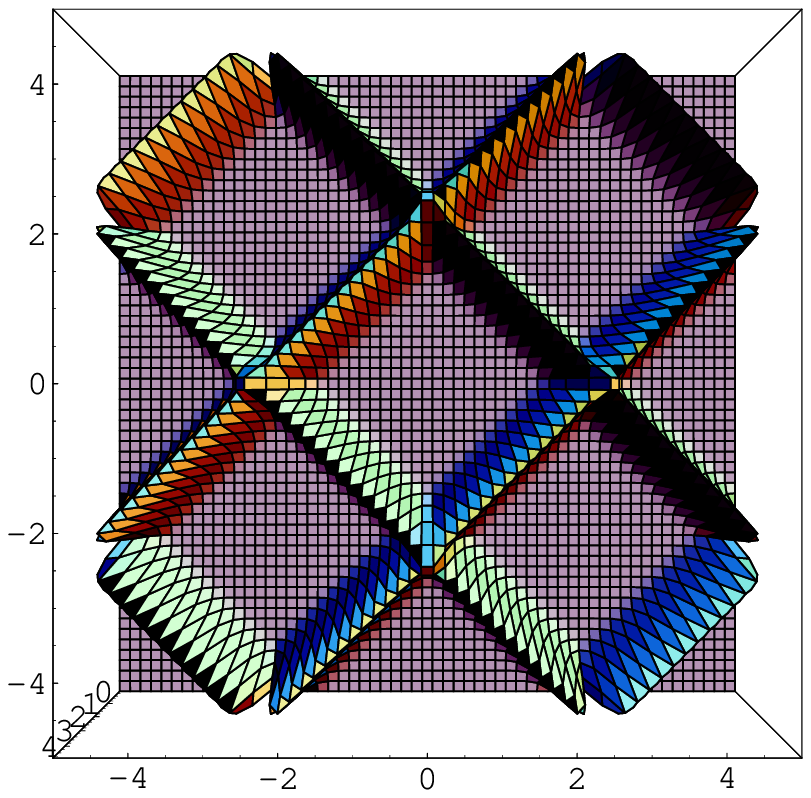}
 &\hspace{0.cm}&
\includegraphics[width=5cm]{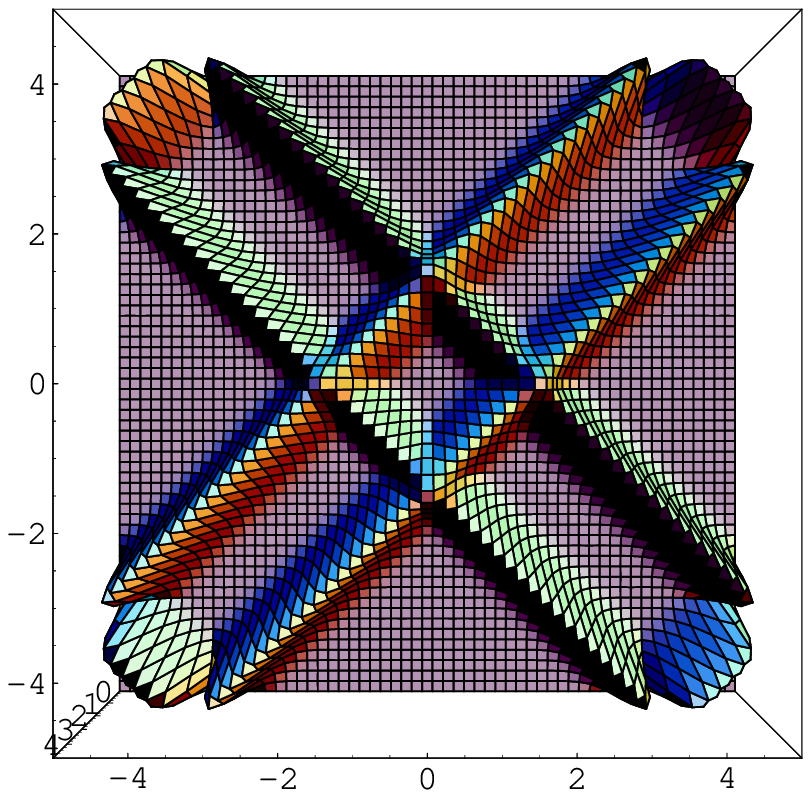}
\end{tabular}}
\caption{Infinite square well: the upper view of the square of the
wave packet $|\psi(u,v)|^2$ for $d=3.5$, $\alpha=10$ (left),
$d=2.5$, $\alpha=10$ (middle), $d=1.5$, $\alpha=5$ (right) and
$L=10$ for all cases.} \label{figwell}
\end{figure}

Now, let us consider the initial behavior of the wave function. The
wave function and the first derivative of the wave function at $v=0$
take the form
\begin{eqnarray}
\Psi(u,0)=\sum_{n=\mbox{\footnotesize{even}}} A_n
\psi_n(u)\psi_n(0),\\
\frac{\partial\Psi(u,v)}{\partial
v}\bigg|_{v=0}=i\sum_{n=\mbox{\footnotesize{odd}}} B_n
\psi_n(u)\psi'_n(0).
\end{eqnarray}
Therefore, the coefficients $A_n$ determine the initial wave
function and the coefficients $B_n$  determine the initial
derivative of the wave function. From a mathematical point of view,
since the underling differential equation (\ref{wd}) is second
order, $A_n$s and $B_n$s are arbitrary and independent variables. On
the other hand, if we are interested to construct wave packets which
simulate the classical behavior with known classical positions and
momentums, all of these coefficients will not be independent. It is
obvious that the presence of odd terms dose not have any effect on
the form of the initial wave function but they are responsible for
the slope of the wave function at $v=0$, and vice versa for the even
terms. Now, consider the behavior of the initial wave function. Near
$v=0$ the differential equation (\ref{wd}) takes the form
\begin{eqnarray}
\left\{-\frac{\partial^2}{\partial u^2}+\frac{\partial^2} {\partial
v^2}+ V(u)-V(0)\right\}\psi(u,v)=0, \label{eq10nearv0}
\end{eqnarray}
which has the solution as
\begin{equation}\label{psi-separated}
\psi(u,v)=\psi(u)\chi(v),
\end{equation}
where $\psi(u)$ and $\chi(v)$ satisfy
\begin{eqnarray}
\frac{d^2\chi_n(v)}{d v^2}+E_n\chi_n(v)&=&0,
\label{eqseparated1}\\
\hspace{-0.6cm}-\frac{d^2\psi_n(u)}{d
u^2}+(V(u)-V(0))\psi_n(u)&=&E_n\psi_n(u),\label{eqseparated2}
\end{eqnarray}
and $E_n$'s are separation constants. These equations are
Schr\"{o}dinger equations with $E_n$'s as their energy levels.
Equation (\ref{eqseparated1}) is exactly solvable with plane wave
solutions
\begin{equation}\label{eqplanewave}
\chi_n(v)=\alpha_n\cos\left(\sqrt{E_n}\,\,v\right)+i\beta_n\sin\left(\sqrt{E_n}\,\,v\right),
\end{equation}
where $\alpha_n$ and $\beta_n$ are arbitrary complex numbers. We can
find the eigenfunctions and the eigenvalues of equation
(\ref{eqseparated2}) using numerical techniques like Spectral Method
\cite{SP} with an acceptable accuracy. Now, the general
solution to Eq.~(\ref{eq10nearv0}) can be written as
\begin{eqnarray}\label{psi-separated2}
\psi(u,v)&=&\sum_{n=\mbox{\footnotesize{even}}} A^*_n
\cos(\sqrt{E_n}v)
\psi_n(u)+i\sum_{n=\mbox{\footnotesize{odd}}}B^*_n\sin(\sqrt{E_n}v)
\psi_n(u).
\end{eqnarray}
As stated before, this solution is valid only for small $v$.
Therefore, we can write the initial conditions as follows
\begin{eqnarray}
\psi(u,0)&=&\sum_{even}A^*_n\psi_n(u),\label{eqinitial1}\\
\psi'(u,0)&=&i\sum_{odd}B^*_n\sqrt{E_n}\psi_n(u),\label{eqinitial2}
\end{eqnarray}
where prime denotes the derivative with respect to $v$. Obviously, a
complete description of the problem would include the specification
of both of these quantities. However, since we are interested to
construct wave packet with classical properties, we need to assume a
specific relationship between these coefficients. The prescription
is that the coefficients have the same functional form
\cite{pedram,pedramPLB3,pedramJCAP} \textit{i.e.}
\begin{eqnarray}\label{eqcanonicalslope}
 A^*_n&=&C(n)\,\,\,\,\,\, \mbox{for $n$
  even},\\
 B^*_n&=&C(n)\,\,\,\,\,\, \mbox{for $n$
  odd},
\end{eqnarray}
where $C(n)$ is a function of $n$. In terms of $A_n$s and $B_n$s we
have
\begin{eqnarray}\label{eqcanonicalslope2}
A_n&=&\frac{1}{\psi_n(0)}C(n)\,\,\,\,\,\,\,\quad \mbox{for $n$
  even},\\
 B_n&=&\frac{\sqrt{E_n}}{\psi'_n(0)}C(n)\,\,\,\,\,\,\, \quad\mbox{for $n$
  odd}.
\end{eqnarray}
First, let us apply this method to the simplest (bounded) case which
is an infinite square well
\begin{equation}
V(q)= \left\{
\begin{array}{ll}
0\qquad\qquad -\frac{L}{2}\leq q\leq \frac{L}{2},\\ \\
\infty\qquad\qquad \mbox{otherwise},
   \end{array}\displaystyle
   \right.
\end{equation}
with well-known orthonormal even and odd eigenfunctions (here $q$
stands for $u$ or $v$)
\begin{equation}
\psi_n(q)= \left\{
\begin{array}{ll}
\sqrt{\frac{2}{L}}\cos(\frac{n \pi q}{L})\qquad\qquad n=1,3,5,...,\\ \\
\sqrt{\frac{2}{L}}\sin(\frac{n \pi q}{L})\qquad\qquad n=2,4,6,....
   \end{array}\displaystyle
   \right.
\end{equation}
Since the energy spectrum for this model is $E_n=\frac{\displaystyle
n^2\pi^2}{\displaystyle L^2}$, using the exact form of the
eigenstates, we can find the desired expansion coefficients
(\ref{eqcanonicalslope2})
\begin{eqnarray}
A_n&=&\sqrt{\frac{L}{2}}C(n)\,\,\,\,\,\,\quad \mbox{for
  even terms},\\
 B_n&=&\sqrt{\frac{L}{2}}C(n)\,\,\,\,\,\,\quad \mbox{for
  odd terms},
\end{eqnarray}
which result in the following wave packet
\begin{equation}
\Psi(u,v)=\sqrt{\frac{2}{L}}\left\{\sum_{n=1,3,5,...}
C(n)\cos(\frac{n \pi u}{L})\cos(\frac{n \pi
v}{L})+i\sum_{n=2,4,6,...} C(n) \sin(\frac{n \pi u}{L})\sin(\frac{n
\pi v}{L})\right\}.
\end{equation}
To completely determine the wave packet, we need to specify the
functional form of $C(n)$. On the other hand, we can choose $C(n)$
by imposing an appropriate initial condition
(\ref{eqcanonicalslope}) which we choose as two Gaussian peaked at
$u=\pm d$
\begin{equation}
\Psi(u,0)=e^{-\alpha\left( u-d \right)^2} +e^{-\alpha\left( u+d
\right)^2 }.
\end{equation}
This choice of initial condition is equal to the following form of
the expansion coefficient
\begin{eqnarray}\nonumber
C(n)&=&\frac{-ie^{\frac{n\pi \left( n\pi  +  4i \alpha d L \right)
}{4\alpha L^2 }}}{{\sqrt{2\alpha L/\pi }}}\left[-
\mbox{erfi}(\frac{n\pi  +
2i\alpha L \left( d - L \right) }{2{\sqrt{\alpha }}L})\right.\\
\nonumber &+&\left. e^{\frac{2i dn\pi }{L}}
\left(\mbox{erfi}(\frac{n\pi  - 2i \alpha L\left( d - L \right)
}{2{\sqrt{\alpha }}L}) -
  \mbox{erfi}(\frac{n\pi  - 2i \alpha L\left( d + L \right)}{2{\sqrt{\alpha
  }}L})\right)\right.\\&+&\left.
  \mbox{erfi}(\frac{n\pi  + 2i \alpha L\left( d + L \right)
 }{2{\sqrt{\alpha }}L})\right],
\end{eqnarray}
where $\mbox{erfi}(q)$ is the imaginary error function
$\mbox{erfi}(q)=-i\mbox{erf}(iq)$. In Fig.~\ref{figwell}, we have
shown the resultant wave packets for different values of $d$ and
$\alpha$. Classically, the particle is free inside the well and can
begin its one-dimensional motion with positive or negative initial
velocity and from any position between $\pm L/2$. As it can seen
from the figure, the behavior of the wave packets are in complete
agreement with the classical scenario and they strongly peak on the
classical trajectories. In fact, the value of $d$ determines the
initial classical position. The presence of two rectangles with
opposite direction shows the two possible direction of initial
motion at $u=\pm d$. Since the wave packets follow some straight
lines, its classical picture consists of a free particle with
constant velocity. Moreover, the height of the crest of the wave
packet is constant along the classical trajectories which shows that
the probability of finding the particle is constant along the
classical path. On the other hand, since the classical velocity of
the particle for this case is a constant of motion, the classical
probability of finding the particle is also constant along its
trajectory. Although this method is generally applicable for
one-dimensional systems, we can also obtain a class of
two-dimensional classical and quantum correspondences for this
especial case. As the figure shows, these wave packets correspond to
the cases of equal absolute velocities in $u$ and $v$ directions
($|v_u|=|v_v|$) with arbitrary initial position.

Now, we study other cases with polynomial or exponential form of the
potential. First, we should choose an appropriate functional form of
$C(n)$. This means that, we need to specify $C(n)$ in such a way
that the initial wave function has a desired classical description
(see Figs.~\ref{fig1}-\ref{fig5}). We will see that this choice of
coefficients leads to complete classical and quantum correspondence
in all the studied cases. In the left part of
Figs.~\ref{fig1}-\ref{fig5}, we have depicted the wave packets for
some sample potentials. As it can be seen from the figures, the wave
packets closely follow the classical paths and peak on them. Before
discussing the solutions, we outline the causal interpretation of
quantum mechanics and use it to interpret the results in the next
section.

\section{The causal interpretation}\label{sec3}
\begin{figure}
\centerline{\begin{tabular}{ccc}
\includegraphics[width=8cm]{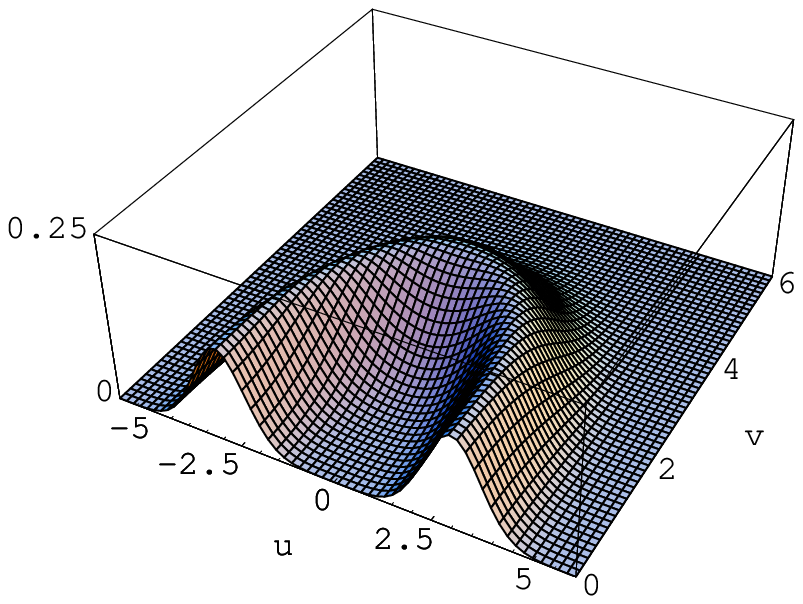}
 &\hspace{2.cm}&
\includegraphics[width=8cm]{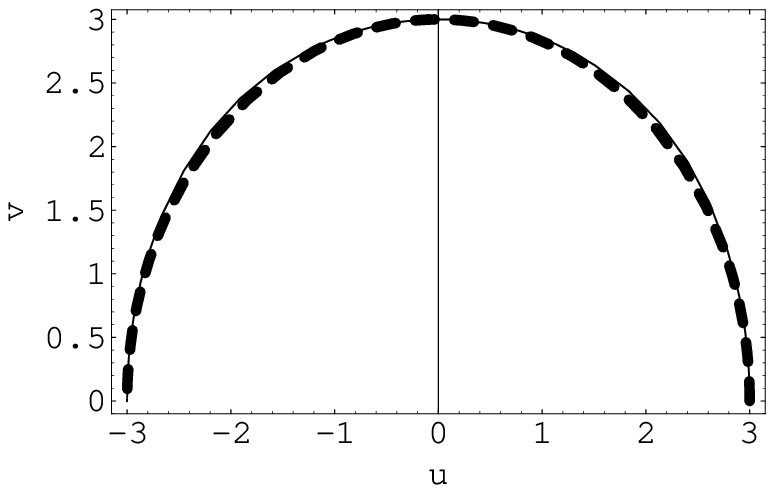}
\end{tabular}}
\caption{Simple Harmonic Oscillator: Left, the square of the wave
packet $| \psi(u,v)|^2$ for
$C(n)=\frac{\,\zeta^n}{\,{\sqrt{2^n\,n!}}}e^{-\zeta^2/4}$ and
$\zeta=3$. Right, the classical (dashed line) and Bohmian (solid
line) trajectories.} \label{fig1}
\end{figure}

\begin{figure}
\centerline{\begin{tabular}{ccc}
\includegraphics[width=8cm]{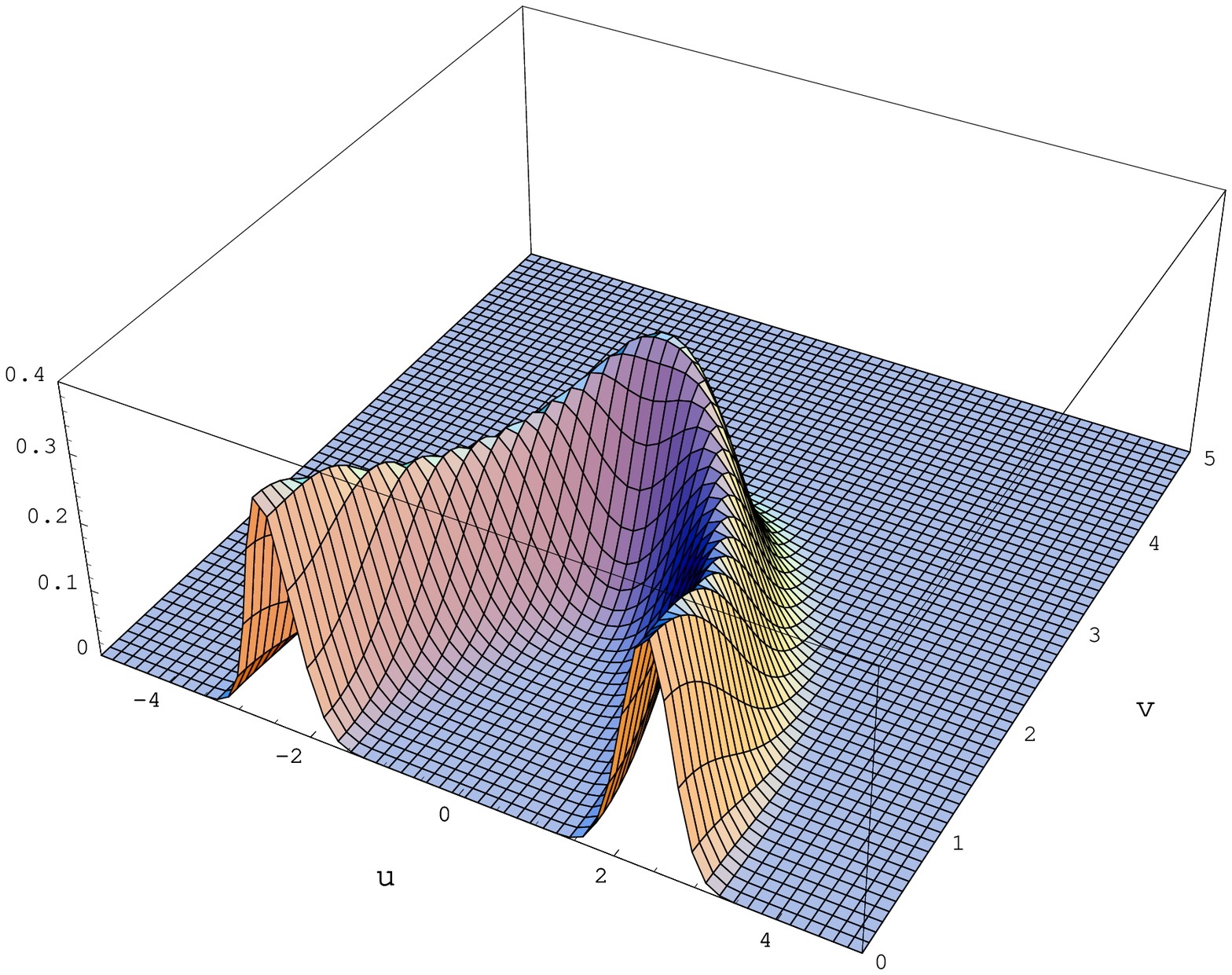}
 &\hspace{2.cm}&
\includegraphics[width=8cm]{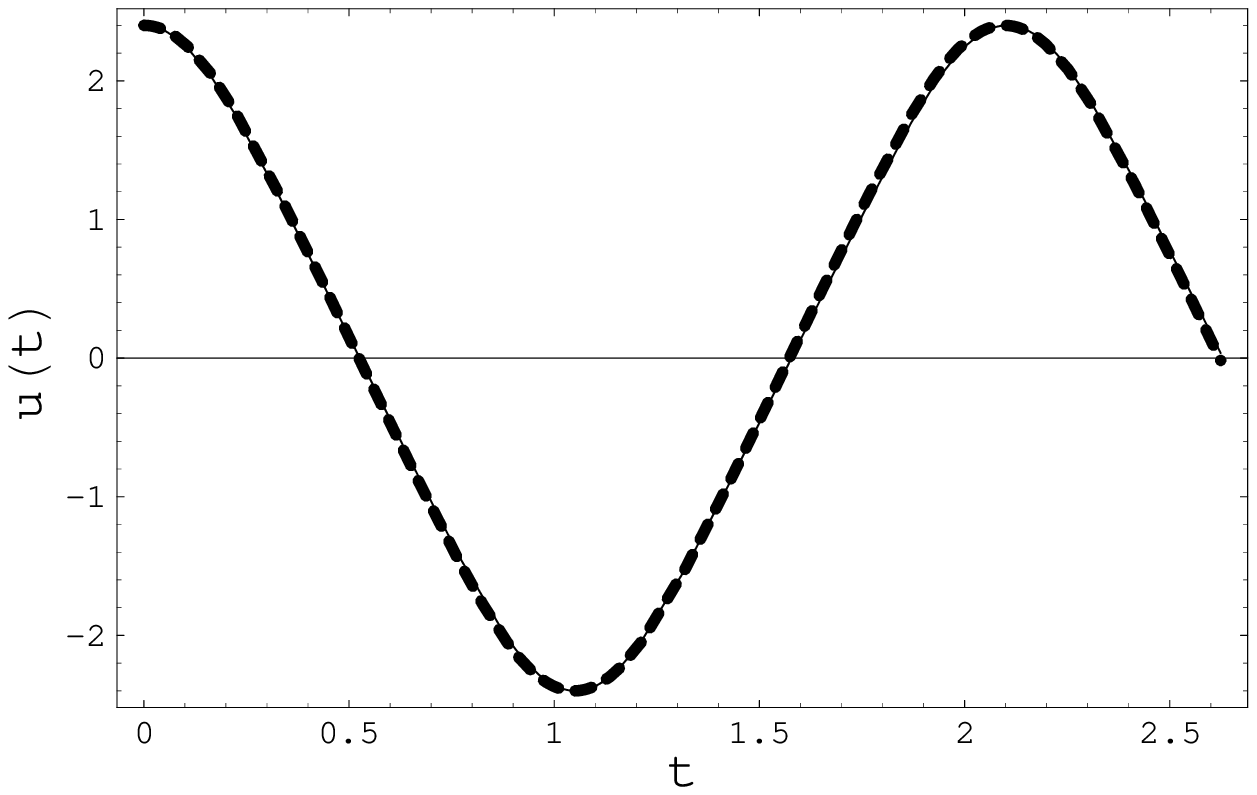}
\end{tabular}}
\caption{$V_2(q)=q^4$: Left, the square of the wave packet $|
\psi(u,v)|^2$ for
$C(n)=\frac{\zeta^n}{\,{\sqrt{2^n\,n!}}}e^{-\zeta^2/4}$ and
$\zeta=4$. Right, the classical (dashed line) and Bohmian (solid
line) trajectories.} \label{fig2}
\end{figure}

\begin{figure}
\centerline{\begin{tabular}{ccc}
\includegraphics[width=8cm]{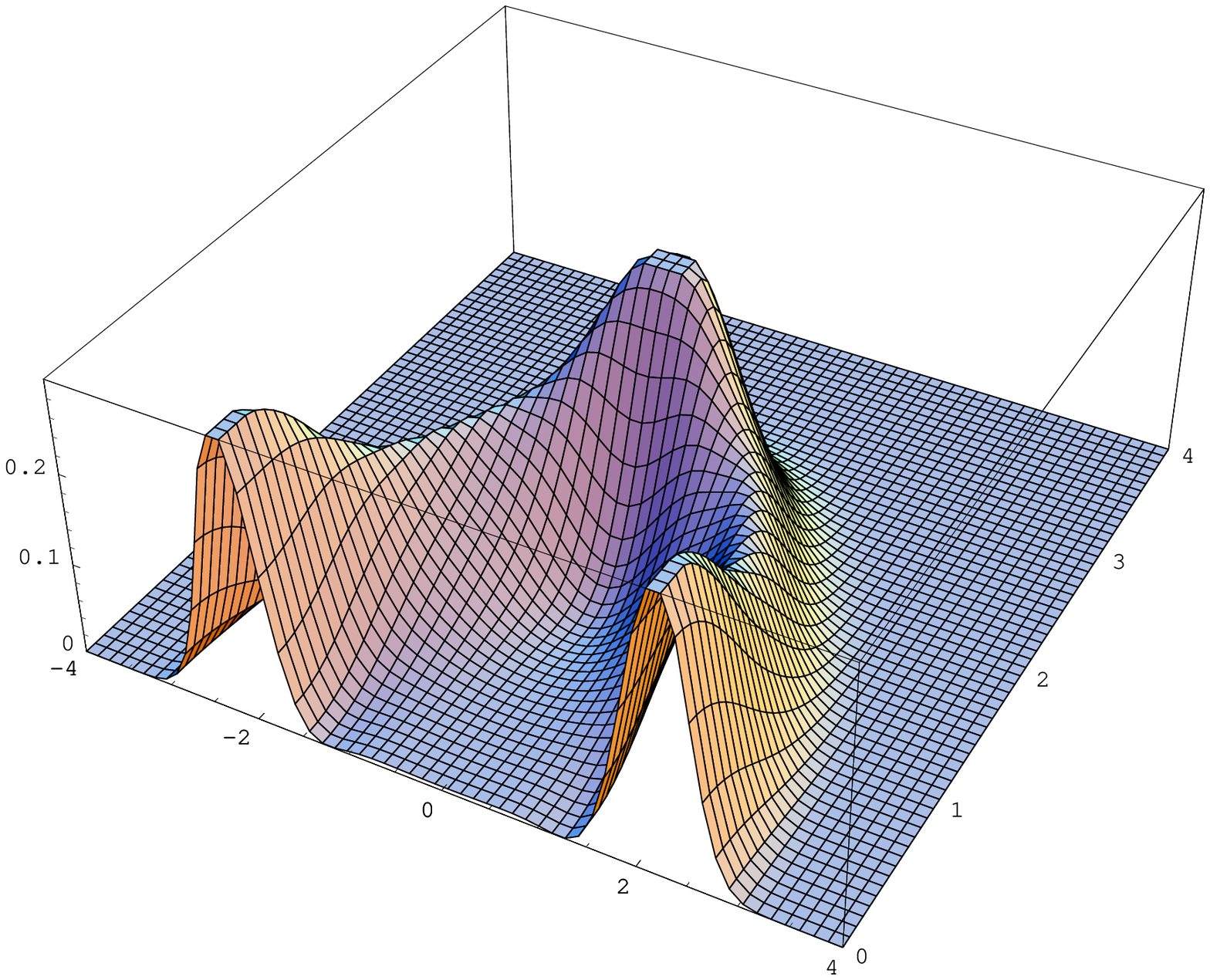}
 &\hspace{2.cm}&
\includegraphics[width=8cm]{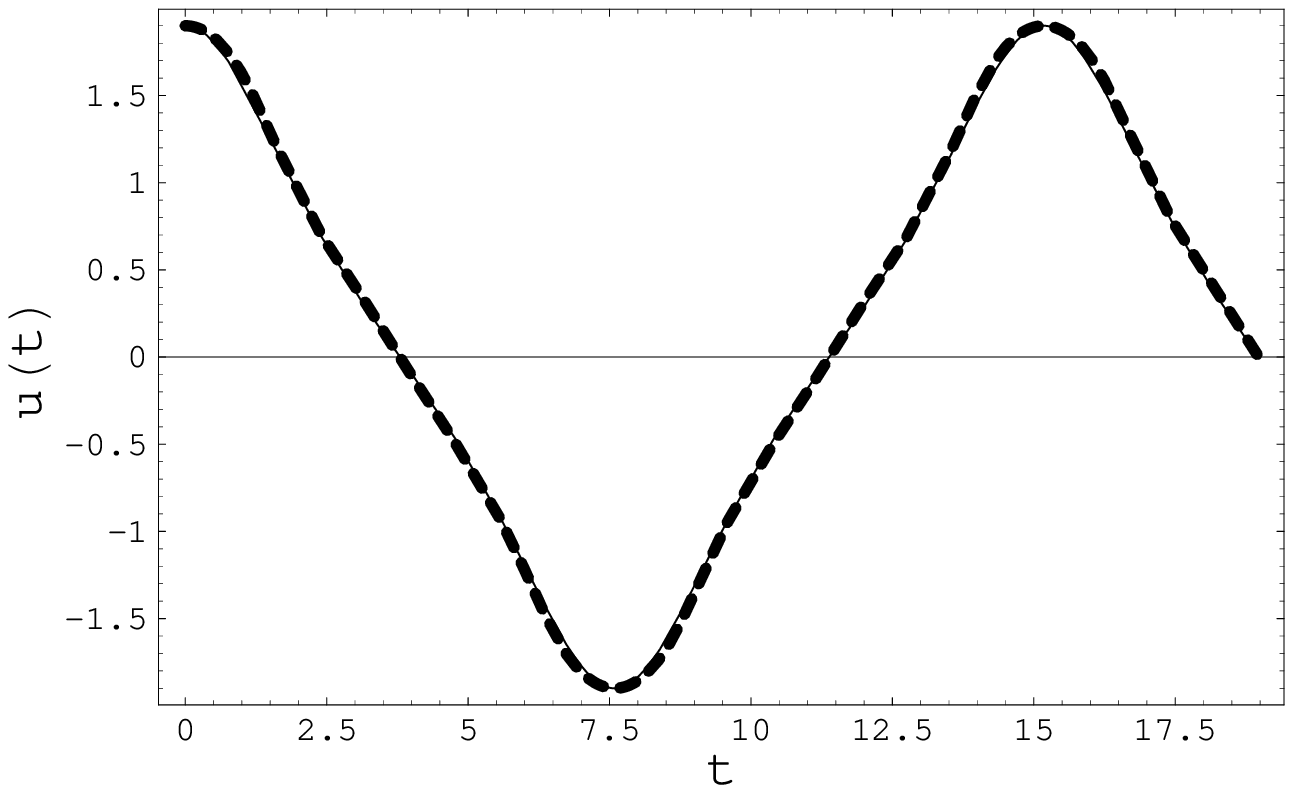}
\end{tabular}}
\caption{Double-Well potential: Left, the square of the wave packet
$| \psi(u,v)|^2$ for
$C(n)=\frac{\,\zeta^n}{\,{\sqrt{2^n\,n!}}}e^{-\zeta^2/4}$ and
$\zeta=3$. Right, the classical (dashed line) and Bohmian (solid
line) trajectories.} \label{fig3}
\end{figure}

\begin{figure}
\centerline{\begin{tabular}{ccc}
\includegraphics[width=8cm]{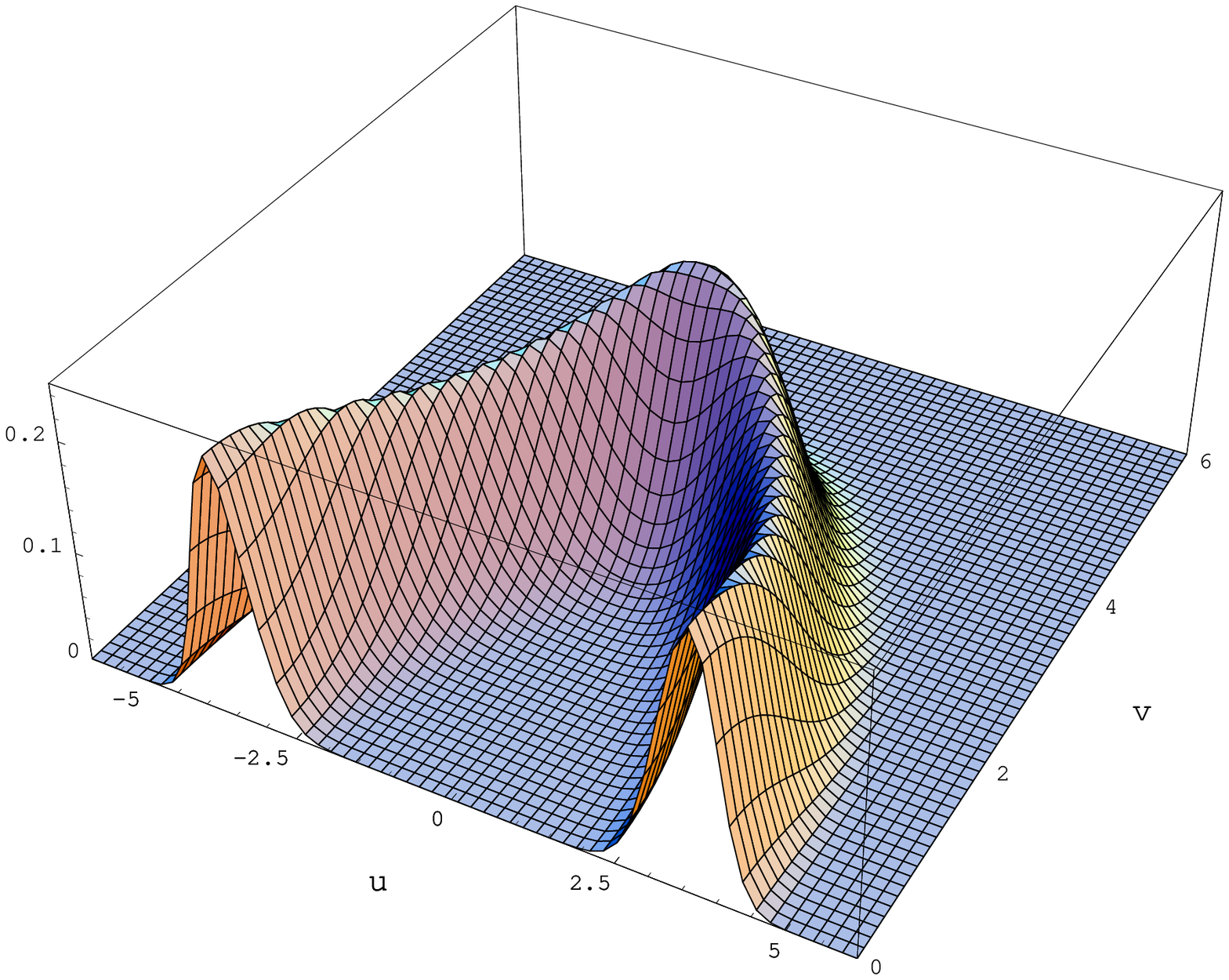}
 &\hspace{2.cm}&
\includegraphics[width=8cm]{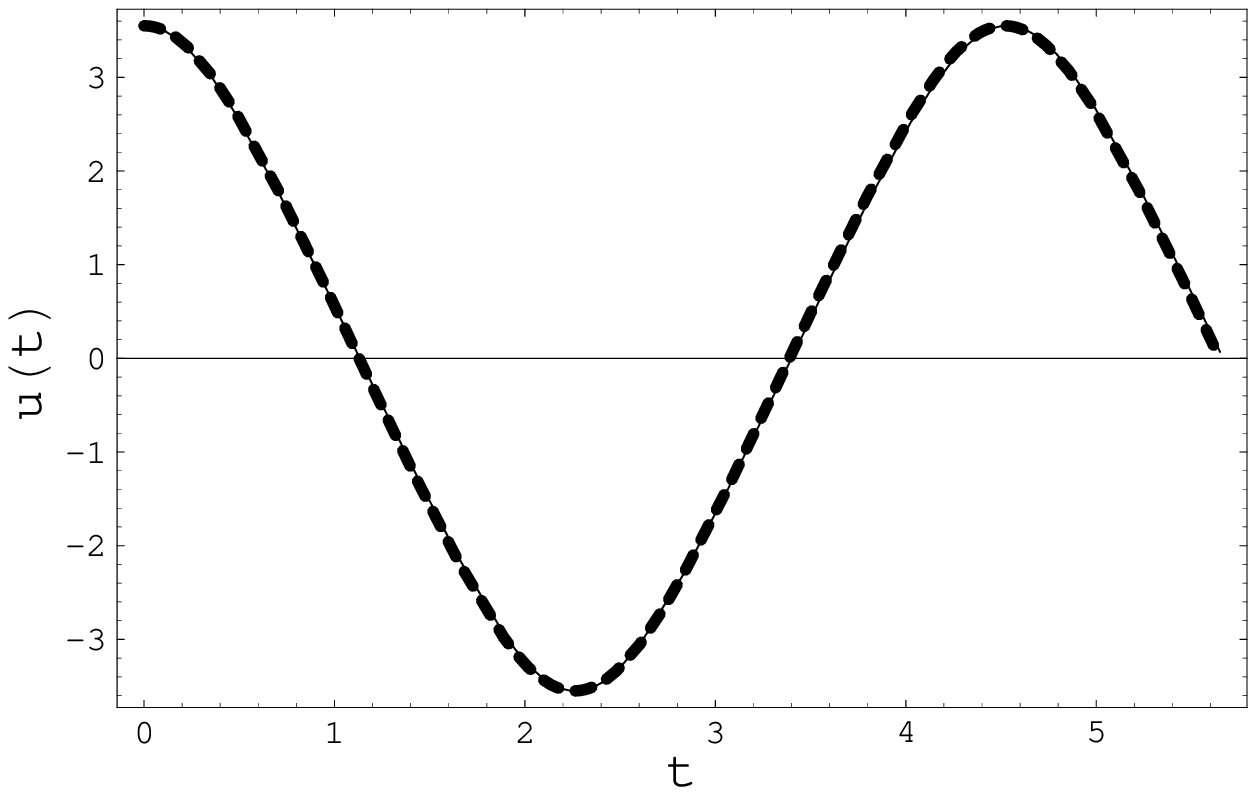}
\end{tabular}}
\caption{$V_4(q)=\exp(q^2/8)-1$: Left, the square of the wave packet
$| \psi(u,v)|^2$ for
$C(n)=\frac{n\,\zeta^n}{\,{\sqrt{2^n\,n!}}}e^{-\zeta^2/4}$ and
$\zeta=3$. Right, the classical (dashed line) and Bohmian (solid
line) trajectories.} \label{fig4}
\end{figure}

\begin{figure}
\centerline{\begin{tabular}{ccc}
\includegraphics[width=8cm]{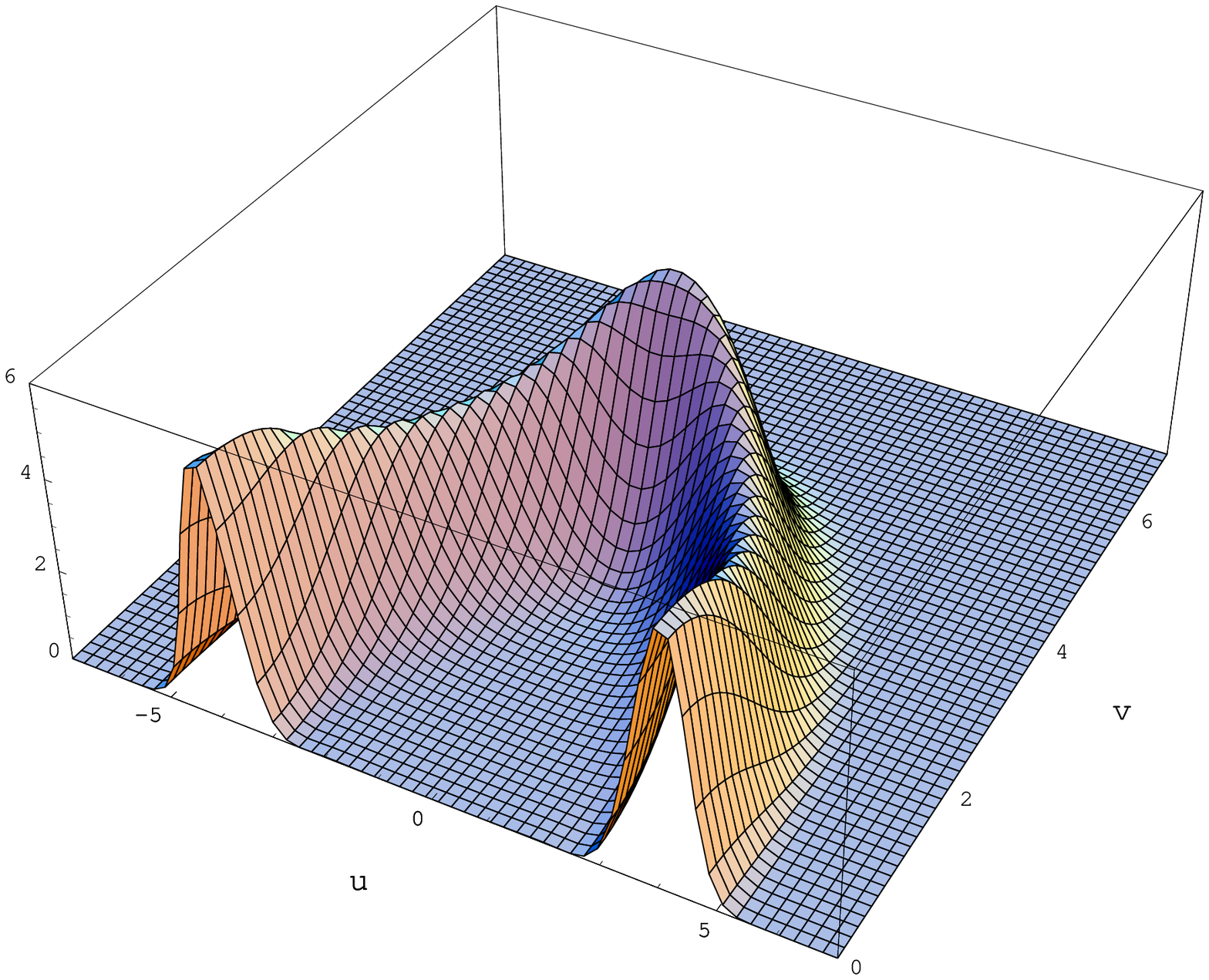}
 &\hspace{2.cm}&
\includegraphics[width=8cm]{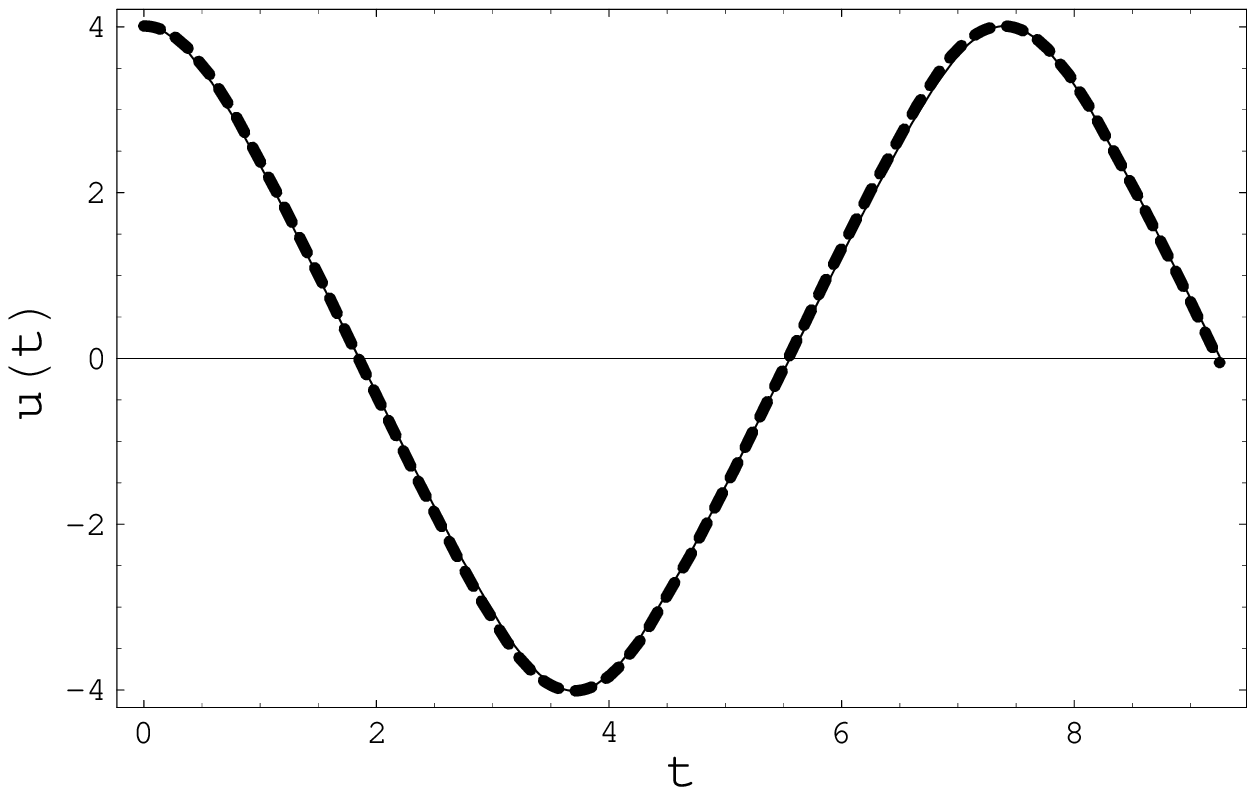}
\end{tabular}}
\caption{$V_5(q)=\cosh(q)-1$: Left, the square of the wave packet $|
\psi(u,v)|^2$ for
$C(n)=\frac{\,\zeta^n}{\,{\sqrt{2^n\,n!}}}e^{-\zeta^2/4}$ and
$\zeta=4$. Right, the classical (dashed line) and Bohmian (solid
line) trajectories.} \label{fig5}
\end{figure}

To make the connection between the classical and quantum results
more concrete, we use the causal interpretation of quantum mechanics
\cite{pedramPLB3,pedramJCAP,holland}. In this interpretation the
wave function can be written as
\begin{equation}\label{R,S}
\Psi(u,v) = R\, e^{iS},
\end{equation}
where $R=R(u,v)$ and $S=S(u,v)$ are real functions and satisfy the
following equations
\begin{eqnarray}
\label{hje} -\frac{\partial^2 R}{\partial u^2} +\frac{\partial^2
R}{\partial v^2}+R\left(\frac{\partial S}{\partial u}\right)^2-
R\left(\frac{\partial S}{\partial v}\right)^2+ (V(u)-V(v))R&=& 0,\\
R\frac{\partial^2 S}{\partial u^2} -R\frac{\partial^2 S}{\partial
v^2}+2 \frac{\partial R}{\partial u} \frac{\partial S}{\partial
u}-2\frac{\partial R}{\partial v} \frac{\partial S}{\partial v}&=&
0.
\end{eqnarray}
To write $R$ and $S$, it is more appropriate to separate the real
and imaginary parts of the wave function
\begin{equation}
\Psi(u,v)=x(u,v)+iy(u,v),
\end{equation}
where $x,y$ are real functions of $u$ and $v$. Using equation
(\ref{R,S}) we have
\begin{eqnarray}
R&=&\sqrt{x^2+y^2},\\
S&=&\arctan(\frac{y}{x}).
\end{eqnarray}
On the other hand, the Bohmian trajectories are governed by
\begin{eqnarray}
p_u = \frac{\partial S}{\partial u},\\
p_v = \frac{\partial S}{\partial v},
\end{eqnarray}
where the momenta correspond to the classical related Lagrangian
$L(q)=\dot{q}^2-V(q)$. Therefore, the Bohmian equations of motion
take the form
\begin{eqnarray}
\dot{u}= \frac{1}{2}\frac{1}{1+\left(\frac{y}{x}\right)^2}\frac{d}{du}\left(\frac{y}{x}\right),\\
\dot{v}=-\frac{1}{2}\frac{1}{1+\left(\frac{y}{x}\right)^2}\frac{d}{dv}\left(\frac{y}{x}\right).
\end{eqnarray}
These differential equations can be solved numerically to find the
time evolution of $u$ and $v$. In the right part of
Figs.~\ref{fig1}-\ref{fig5}, we have shown the Bohmian trajectories
as solid lines for five different forms of the potentials
\begin{eqnarray}
\left\{
  \begin{array}{ll}
    V_1(q)=q^2, &  \\
    V_2(q)=q^4, &  \\
    V_3(q)=-3q^2+q^4, & \\
    V_4(q)=\exp(q^2/8)-1, & \\
    V_5(q)=\cosh(q)-1. &
  \end{array}
\right.
\end{eqnarray}
The initial wave functions are chosen to be localized on classical
initial positions and also the initial slope is determined by
equations (\ref{eqinitial1}) and (\ref{eqcanonicalslope}). In fact,
we are free to choose any arbitrary but appropriate initial wave
functions which correspond to the classical scenario. For cases
studied here, we choose two different forms of coefficients
(Figs.~\ref{fig1}-\ref{fig5})
\begin{eqnarray}
C(n)=\frac{\zeta^n}{\,{\sqrt{2^n\,n!}}}e^{-\zeta^2/4},\hspace{.5cm}
C(n)=\frac{n\,\zeta^n}{\,{\sqrt{2^n\,n!}}}e^{-\zeta^2/4}.
\end{eqnarray}
These coefficients are chosen so that the initial wave function
contains two bumps along $u-$axis which correspond to initial and
final classical positions. In fact, these choices are not special
and there are many other equivalent alternatives. Since the slope of
the wave function is related to the odd terms, the expectation value
of the momentum operator is positive and negative respectively for
initial and final states of motion in agreement with the classical
scenario. Note that, since these two bumps are related to two
different physical situations, the integrations are along positive
and negative parts of the $u-$axis, respectively. Figure \ref{fig1}
shows the resulting wave packet for the simple harmonic oscillator
and the Bohmian and classical trajectories. In the right part of
figures~\ref{fig2}-\ref{fig5}, we have also shown the Bohmian and
classical behaviors of variable $u$ versus time ($u(t)$). The
complete correspondence between these results are again manifest. It
is worth to mention that for the cases where $\dot{u}(0) \ne 0$, we
need to choose complex $C(n)$ \cite{pedramPLB4}. Beside the stated
advantages, this method has also some limitations. As we have shown
by various examples, this method works well for symmetric bounded
potentials. This class of potentials results in oscillatory
classical motions which can be followed by wave packets with a
finite set of eigenfunctions. Therefore, this method, in the present
fashion, is not applicable even for simple unbounded situations like
free particle, potential barrier, delta function potential and etc.
The generalization of this method for the case of free particles is
the subject of our future work.

\section{Conclusions}\label{sec4}
In this work, we have studied the classical-quantum correspondence
in the context of non-relativistic quantum mechanics. First, we
considered two similar oscillators for $u$ and $v$ variables and
obtained the classical trajectories. Since these two variables
satisfy the same equation of motion, we obtained the trajectory of
one variable versus another one by choosing a particular set of
initial condition. In the quantum version, this scenario leads to a
hyperbolic partial differential equation where its solution contains
infinite unknown coefficients. Half number of these coefficients are
related to the initial form of the wave function and others
correspond to the initial slope of the wave function. Upon using a
specific relation between these coefficients and choosing
appropriate initial wave functions we can construct desired wave
packets. The crests of these wave packets closely follow the
classical paths from initial position to the final position. We have
quantified this correspondence using de-Broglie Bohm interpretation
of quantum mechanics. We applied this method to various cases and
showed that the resulting wave packets completely simulate their
classical counterpart's behavior. In particular, the Bohmian
trajectories coincided well with classical trajectories for the all
cases.

\section*{Acknowledgements}
I would like to thank M. Mirzaei for his useful comments and
discussions.

\end{document}